\documentclass[%
 preprint,
 amsmath,amssymb,
 aps,
superscriptaddress]{revtex4-2}

\usepackage{graphicx}
\usepackage{dcolumn}
\usepackage{bm}

\DeclareUnicodeCharacter{2212}{-}
\begin{document}


\title{Charge density wave-templated spin cycloid in topological semimetal NdSb$_{x}$Te$_{2-x-\delta}$}

\author{Tyger H. Salters}
\affiliation{Department of Chemistry, Princeton University, Princeton, New Jersey 08544, United States}
\author{Fabio Orlandi}
\affiliation{ISIS Neutron Pulsed Facility, Science and Technology Facilities Council, Rutherford Appleton Laboratory, Oxford OX11 0QX, UK}
\author{Tanya Berry}
\affiliation{Department of Chemistry, Princeton University, Princeton, New Jersey 08544, United States}
\author{Jason F. Khoury}
\affiliation{Department of Chemistry, Princeton University, Princeton, New Jersey 08544, United States}
\author{Ethan Whittaker}
\affiliation{Department of Chemistry, Princeton University, Princeton, New Jersey 08544, United States}
\author{Pascal Manuel}
\affiliation{ISIS Neutron Pulsed Facility, Science and Technology Facilities Council, Rutherford Appleton Laboratory, Oxford OX11 0QX, UK}
\author{Leslie M. Schoop}
\email{lschoop@princeton.edu}
\affiliation{Department of Chemistry, Princeton University, Princeton, New Jersey 08544, United States}

\begin{abstract}
Magnetic topological semimetals present open questions regarding the interplay of crystal symmetry, magnetism, band topology, and electron correlations. \textit{Ln}Sb$_{x}$Te$_{2-x-\delta}$ (\textit{Ln}= lanthanide) is a family of square-net-derived topological semimetals that allows compositional control of band filling, and access to different topological states via an evolving charge density wave (CDW) distortion. Previously studied Gd and Ce members containing a CDW have shown complex magnetic phase diagrams, which implied that spins localized on \textit{Ln} interact with the CDW, but to this date,  no magnetic structures have been solved within the CDW regime of this family of compounds. Here, we report on the interplay of the CDW with magnetism in NdSb$_{x}$Te$_{2-x-\delta}$, by comparing the undistorted square net member NdSb$_{0.94}$Te$_{0.92}$ with the CDW-distorted phase NdSb$_{0.48}$Te$_{1.37}$, via single-crystal x-ray diffraction, magnetometry, heat capacity, and neutron powder diffraction. NdSb$_{0.94}$Te$_{0.92}$ is a collinear antiferromagnet with T$_N$ $\sim$ 2.7 K, where spins align antiparallel to each other, but parallel to the square net of the nuclear structure. NdSb$_{0.48}$Te$_{1.37}$ exhibits a nearly five-fold modulated CDW (\textbf{q$_{CDW}$}=0.18), isostructural to other \textit{Ln}Sb$_{x}$Te$_{2-x-\delta}$ at similar \textit{x}. NdSb$_{0.48}$Te$_{1.37}$ displays more complex magnetism with T$_N$ = 2.3 K, additional metamagnetic transitions, and an elliptical cycloid magnetic structure with \textbf{q$_{mag}$}=-0.41\textbf{b$^{\ast}$}. 
The magnitudes of q$_{CDW}$ and q$_{mag}$ exhibit a integer relationship, $1+2q_{mag}=q_{CDW}$, implying a coupling between the CDW and magnetic structure. Given that the CDW is localized within the nonmagnetic distorted square net, we propose that conduction electrons ``template'' the spin modulation via the Ruderman–Kittel–Kasuya–Yosida interaction.
\end{abstract}

\maketitle


\section{Introduction}
In the search for new topological matter, materials with a square-net structural motif have been heavily studied as a popular platform to host topological electronic states.\cite{klemenz2019topological, klemenz2020role, klemenz2020systematic, schoop2018chemical, liu_nearly_2016,park_anisotropic_2011,hu_evidence_2016,wu_nonsymmorphic_2022} The $p_{x}$ and $p_{y}$ orbitals that contribute much of the bonding in the square-net yield band structures with multiple topological band crossings, and thus the materials have been found to host Dirac, Weyl, and nodal line topological semimetals (TSMs), as well as topological insulators\cite{Kefeng2012CaMnBi2,borisenko2019YbMnBi2,schoop2016dirac}. From a chemical standpoint, the bonding within the square net can be described as ``hypervalent'', which means that it is delocalized over the square net, as the electrons occupy an exactly half-filled $p_x$/$p_y$ band. Deviation from a half-filled electron count will disturb this type of bonding, causing electron localization that is accommodated by distortions of the net.\cite{a2000hypervalent} Such a deviation is reflected in the band structure by the appearance of nesting at the Fermi surface, wherein pockets coincide by translation of some wavevector \textit{\textbf{q}}.\cite{gruner_density_1994} Such an instability can be alleviated by formation of a charge density wave (CDW), in which bond localization, also influenced by electron-phonon coupling, leads to a periodic distortion of the structure with a modulation wavevector reflecting the nesting vector \textit{\textbf{q}}.\cite{johannes_fermi_2008}

The canonical nodal line semimetal ZrSiS, as well as several isostructural relatives, contains a nodal line at the Fermi level and a non-symmorphic Dirac node above $E_{F}$.\cite{schoop2016dirac} Members of the family crystallize in the tetragonal \textit{P4/nmm} space group, in which, if using ZrSiS as an example, the Si square net lies in the \textit{ab} plane. The square net is sandwiched by layers of Zr and S along the \textit{c} axis, in a puckered rock-salt layer between each square net. One isostructural subgroup of the ZrSiS family, \textit{Ln}SbTe (\textit{Ln}=Lanthanide), affords chemical tuning of $E_{F}$ by substitution of Te into the Sb square net\cite{lei2021band}, allowing access to the non-symmorphic Dirac node; such solid solutions are largely overlooked by popular databases used to predict TSMs due to the limits of the computational methods which they are built upon.\cite{zunger2019beware, bradlyn2017topological, weiland2019band, vergniory2019complete} Magnetic rare earth ions also make the \textit{Ln}SbTe family attractive to study the interplay of topological states with magnetism. This allows for opportunities to observe new topological states through the breaking of time-reversal symmetry, leveraging magnetic order as another tunable parameter to control electronic structure and access multiple topologically distinct phases.\textit{Ln}Sb$_{x}$Te$_{2-x-\delta}$ ($\delta$ indicating vacancy concentration in the square net) materials also exhibit charge density waves (CDWs), which are coupled with structural distortions in the square-net that evolve continuously with Sb/Te composition, and are a result of changes in band filling within the square net.\cite{dimasi1996LaSbTe, lei2019GdSbTe,singha2021evolving} We have recently reported that the CDW is responsible for the gapping of topologically trivial states in GdSb$_{0.46}$Te$_{1.48}$, yielding an ideal nonsymmorphic Dirac semimetal band structure free of all other states about \textit{E$_{F}$}, separated by hundreds of meV.\cite{lei2021band} Depending of the doping level, the CDW can have one or multiple q-vectors that describe the structure modulation. The single q-vector region usually lies in the intermediate doping range. Here, the q-vector is in-plane and agrees with a nesting vector that connects the two sheets of the diamond-shaped Fermi surface that is typical to such square-net materials.\cite{lei2021band,dimasi1996LaSbTe} GdSb$_{x}$Te$_{2-x-\delta}$ also displays highly complex magnetism in which magnetic phase diagrams evolve with the compositionally driven CDW.\cite{lei2021magphase} In-plane magnetic anisotropy has been reported in mechanically detwinned GdSb$_{0.46}$Te$_{1.48}$ crystals, indicating that the unidirectional CDW distortion yields especially complex H-T magnetic phase diagrams along the direction of the distortion. Notably, magnetoentropic analysis of this composition also suggests the signature of an antiferromagnetic skyrmion phase at low applied field.\cite{lei2021magphase} Across all distorted phases, the CDW does not significantly distort the puckered magnetic \textit{Ln}Te layers in the parent structure. However, the evolving distortion and composition of the nonmagentic (Sb,Te) square net is suggested to influence couplings through Ruderman–Kittel–Kasuya–Yosida (RKKY) interaction. RKKY interactions are sensitive to the also-evolving density of states, yielding a Sb-concentration dependent magnetic phase diagram. 

The highly neutron-absorbing nature of Gd complicates the feasibility of determining the magnetic structures of GdSb$_{x}$Te$_{2-x-\delta}$ with neutron diffraction. In an effort to study evolving magnetism and topology with neutron-friendly analogues, we have previously synthesized isostructural CeSb$_{x}$Te$_{2-x-\delta}$. \cite{singha2021evolving} The stoichiometric parent compound, CeSbTe, is capable of switching between Dirac and Weyl semimetal states by changing the strength of an applied magnetic field.\cite{schoop2018tunable} Single-crystal x-ray diffraction (SCXRD) has shown that CDWs are continuously tunable in CeSb$_{x}$Te$_{2-x-\delta}$ and exhibit modulated structures consistent with GdSb$_{x}$Te$_{2-x-\delta}$ at similar Sb compositions. Distortions and calculations on CeSb$_{0.51}$Te$_{1.40}$ suggest that the approach of driving CDWs to discover ``clean'' TSMs can be extended to several members of the \textit{Ln}Sb$_{x}$Te$_{2-x-\delta}$ family\cite{singha2021evolving}. In the single-q CDW region of the phase diagram, the magnetism of CeSb$_{x}$Te$_{2-x-\delta}$ does not exhibit the same complexity, but does yield a composition and distortion dependent magnetic phase diagram. Spins align out-of-plane in CeSb$_{x}$Te$_{2-x-\delta}$, perpendicular to the square net, which is consistent with neutron diffraction studies of the parent compound, CeSbTe,\cite{schoop2018tunable} and divergent from GdSb$_{x}$Te$_{2-x-\delta}$, which exhibits an in-plane easy axis. Thus, in the Ce system, the spins do not align with the CDW q-vector, at least in the single q-vector region, which might be the cause of its less complex magnetism. We now direct our efforts to a different neutron-friendly analog, NdSb$_{x}$Te$_{2-x-\delta}$, where spins are oriented in-plane, similarly to GdSb$_{x}$Te$_{2-x-\delta}$.

NdSbTe has been reported to exhibit metamagnetic transitions and field-dependent transport, implying an interaction of magnetism with electronic structure \cite{pandey_electronic_2020}. Herein we report the magnetism and magnetic structures of NdSb$_{0.91}$Te$_{0.89}$ and NdSb$_{0.48}$Te$_{1.37}$, two members of NdSb$_{x}$Te$_{2-x-\delta}$, which we studied by magnetometry, single crystal x-ray diffraction, and neutron powder diffraction. NdSb$_{0.91}$Te$_{0.89}$ is found to have a crystal structure consistent with other tetragonal \textit{Ln}Sb$_{x}$Te$_{2-x-\delta}$ phases, and exhibits a collinear magnetic structure in which spins align within the \textit{ab} plane. NdSb$_{0.48}$Te$_{1.37}$ is nearly isostructural to GdSb$_{0.46}$Te$_{1.48}$ with a single-q CDW modulation; and displays a complex elliptical cycloid magnetic structure. The magnitude of the magnetic propagation vector q$_{mag}$ is found to have an integer relationship with that of q$_{CDW}$, $1+2q_{mag}=q_{CDW}$, which implies the interaction of conduction electrons from the distorted square net with the unpaired electrons in Nd that lead to magnetism, which are considered to sit at energy levels far below the Fermi energy. This is the first magnetic structure solved for a CDW \textit{Ln}Sb$_{x}$Te$_{2-x-\delta}$ phase, and may give insight into the interaction between magnetism and topological bands in these ``ideal'' TSMs.

\begin{table}[b]
\caption{Easy-axis and neutron absorptivity in \textit{Ln}Sb$_{x}$Te$_{2-x-\delta}$}
\begin{ruledtabular}
\begin{tabular}{ccc}
\textrm{(\textit{Ln} in \textit{Ln}Sb$_{x}$Te$_{2-x-\delta}$)} &
\textrm{Spin alignment} &
\textrm{Neutron absorption cross-section (barn)\cite{sears_neutron_1992}} \cr
\colrule
Gd & $\rightarrow (\parallel ab)$\cite{lei2021magphase} & 49700  \\

Ce & $\uparrow (\perp ab)$\cite{singha2021evolving} & 0.63  \\

Nd & $\rightarrow (\parallel ab)$$^{this~work}$ & 50.5  \\
\end{tabular}
\end{ruledtabular}
\end{table}

\section{Experimental}

\subsection{Synthesis of NdSb$_{0.91}$Te$_{0.89}$ and NdSb$_{0.48}$Te$_{1.37}$}
Single crystals of NdSb$_{0.91}$Te$_{0.89}$ and NdSb$_{0.48}$Te$_{1.37}$ were prepared using a chemical vapor transport (CVT) method. Prior to CVT growth, precursor powders with nominal compositions of NdSb$_{0.5}$Te$_{1.5}$ and NdSbTe were prepared from Nd pieces (Sigma Aldrich, 99.9\%), Sb shot (Alfa Aesar, 99.999\%), and Te pieces which were purified in-house after purchase from a commercial source (Sigma Aldrich, 99.999\%). The elements were loaded into a 12 mm fused quartz tube and flame sealed under vacuum with an Ar backfill at approximately 60 mTorr, heated to 1000 $^{\circ}$C over 12 hr, held for 2d and cooled over 12 hr in a box furnace. The resulting powders were ground and loaded into a 16 mm tube with 120 mg I$_{2}$ as a vapor transport agent, sealed under vacuum resulting in tubes of 6-8 cm in length. CVT growth was carried out by placing the tubes in a single zone tube furnace, with the powder sides placed in the middle of the furnace to create a thermal gradient at the sink side. The tubes were heated to 1000 $^{\circ}$C over 12 hr, held for 7d and cooled over 12 hr, resulting in crystals formed at the sink end. The elemental composition and approximate stoichiometry
of the structure were confirmed via scanning electron microscopy (SEM)/Energy-Dispersive
Spectroscopy (EDS). In attempted preparation of NdSbTe, Sb loss was frequently observed during CVT growth, confirmed by composition analysis. 

Powder phase samples for neutron diffraction were prepared from the elements using a vapor-assisted technique. Samples were sealed in quartz tubes with I$_{2}$ identically to the previously described procedure. The tubes were heated in a box furnace to 950 $^{\circ}$C over 12 hr, held for 7d and cooled over 12 hr, yielding highly crystalline powders. Single crystals of NdSb$_{0.48}$Te$_{1.37}$, of size and quality for magnetic and specific heat measurement were also obtained from this method in the same tube.

\subsection{Single-crystal X-ray Diffraction and Magnetic Measurements}
SCXRD data for NdSb$_{0.91}$Te$_{0.89}$ and NdSb$_{0.48}$Te$_{1.37}$ were collected on a Bruker D8 VENTURE with a PHOTON 3 CPAD detector and a graphite-monochromatized Mo-$K_{\alpha}$ radiation source.  Integrated data were corrected with a multiscan absorption correction
using SADABS. The structure of NdSb$_{0.91}$Te$_{0.89}$ was solved via SHELXT\cite{sheldrick2015shelxt} using intrinsic phasing and refined
with SHELXL\cite{sheldrick_short_2008} using the least squares method, as implemented in the OLEX2 program.\cite{dolomanov_olex2_2009} Satellite peaks, indicative of a structural modulation were observed in NdSb$_{0.48}$Te$_{1.37}$. The satellite peaks were indexed to an incommensurate modulation of \textit{\textbf{q}}=0.18\textit{\textbf{b}}$^{\ast}$. Refinement was thus carried out using the superspace approach, where the displacive distortion of atomic positions is expressed by a periodic modulation function, yielding a 3 + 1 dimensional superspace group. Refinements with the superspace approach were carried out in JANA2006.\cite{Petricek2014} This approach included positional, ADP and site occupancy modulations in the refinement. Determination of Sb/Te occupancy and ordering within the square-net was limited by near-identical scattering power due to their closeness in atomic number (Z = 51 and 52, respectively). At laboratory-accessible X-ray wavelengths, the two species were indistinguishable if present on the same crystallographic site. Thus, for all refinements, atomic occupancies within the Sb/Te square-net were constrained to the stoichiometry derived from EDS.

Magnetic measurements were conducted using the vibrating sample magnetometer (VSM) option of either a Quantum Design DynaCool Physical Property Measurement System (PPMS)
or a Magnetic Property Measurement System (MPMS) 3 SQUID magnetometer between 1.8 and 300 K. Single crystals were fixed between quartz paddles in a brass sample holding using GE Varnish, with field applied parallel and perpendicular to the \textit{ab} plane. Magnetic phase diagrams were generated by collecting temperature-dependent magnetization curves during warming, with varying fields, up to 0.1 T in 0.01 T increments, then up to 5.0 T in 0.1 T increments. Critical temperatures and phase boundaries are derived from peaks in the temperature derivative of the DC magnetic susceptibility, d$\chi$/dT. d$\chi$/dT curves were plotted to build the magnetic phase diagram in the \textit{H-T} space. 

\subsection{Heat Capacity}
Heat Capacity (C$_{P}$) measurements of NdSb$_{0.48}$Te$_{1.37}$ were carried out using a Quantum Design DynaCool PPMS using the semiadiabatic method. The single-crystal sample was fixed to the stage using Apiezon N grease, and measurements were carried out between 1.8 and 300 K. Variable-field measurements were conducted by applying a magnetic field perpendicular to the \textit{ab} plane.
\subsection{Neutron powder diffraction}
Neutron powder diffraction data were collected on the WISH instrument at the ISIS Muon and Neutron Source (UK). \cite{Chapon2011} The samples were contained in sealed Vanadium cans and the neutron diffraction data were collected in the 1.5 -- 5 K temperature range. The diffraction data were collected in five detectors banks each covering 32 degrees of the scattering plane. The Rietveld refinement of the powder data were conducted with the help of the JANA2020 software. \cite{Petricek2014, petricek_jana2020_2021} Group theoretical calculations and symmetry analysis were performed with the help of the ISOTROPY \cite{Stokes_ISO} and ISODISTORT\cite{Campbell2006} software and of the Bilbao crystallographic server \cite{Aroyo2006A,Aroyo2006B,Elcoro2017}.

\section{Results and discussion}

\subsection{Crystal Structure}
\begin{figure}[hbt!]
  \includegraphics[width=\textwidth]{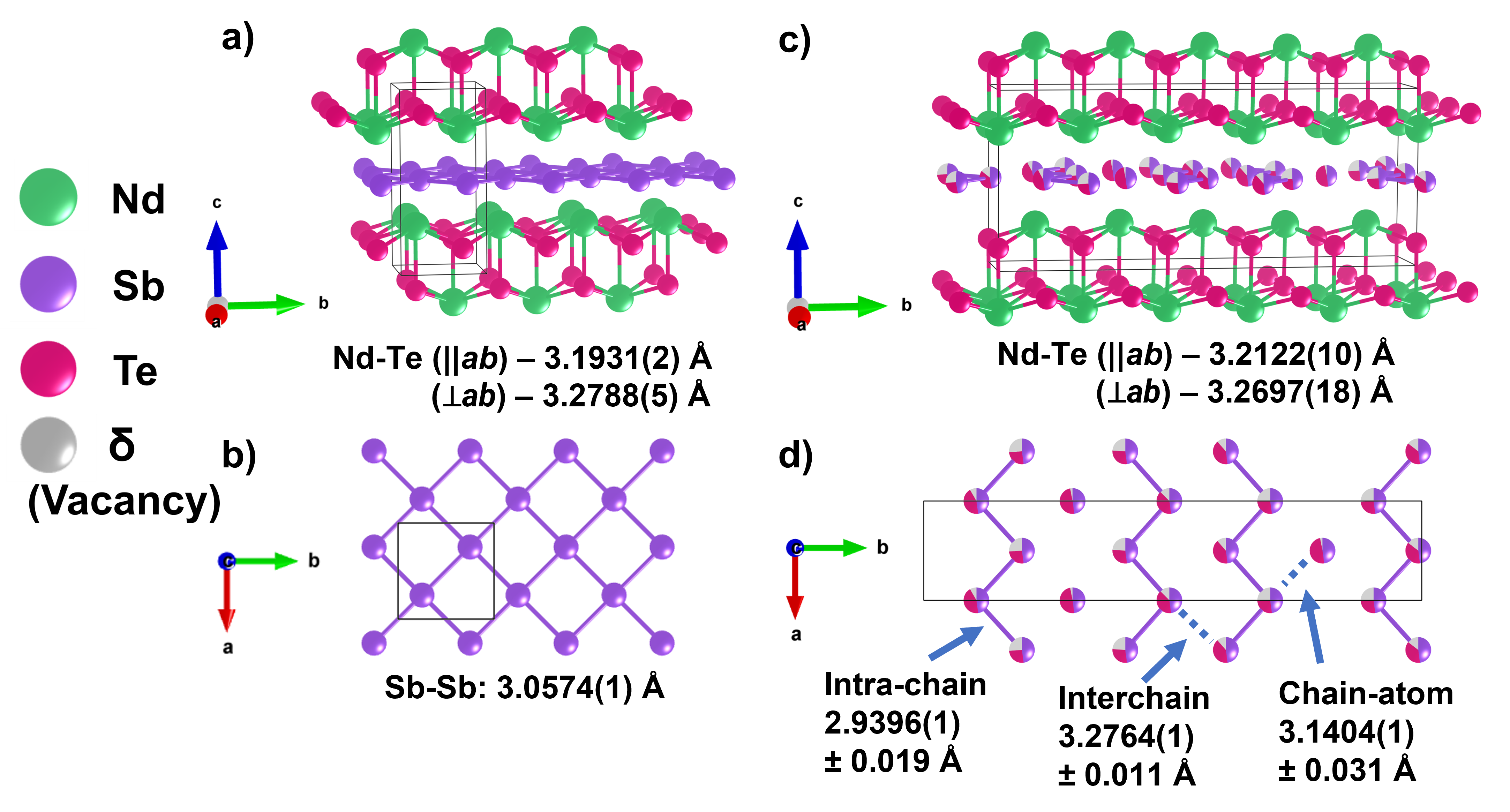}
  \caption{Crystal structures of (a,b) NdSb$_{0.91}$Te$_{0.89}$ and (c, d) NdSb$_{0.48}$Te$_{1.37}$, as determined by single-crystal x-ray diffraction with bond distances given. Bonds in the distorted net of NdSb$_{0.48}$Te$_{1.37}$ are drawn for distances equal to or shorter than the determined Sb-Sb bond length in undistorted NdSb$_{0.91}$Te$_{0.89}$, emphasizing the loss of bond delocalization upon CDW. Bond distances in d) are shown as their average lengths $\pm$ distance modulation, with errors given in parenthesis. }
  \label{structure}
  \centering
\end{figure}

The crystal structures of NdSb$_{0.91}$Te$_{0.89}$  and NdSb$_{0.48}$Te$_{1.37}$, as determined by SCXRD at 100 K are shown in Figure 1. The structure and lattice parameters of NdSb$_{0.91}$Te$_{0.89}$ are consistent with other reported tetragonal \textit{Ln}Sb$_{x}$Te$_{2-x-\delta}$ members, where no CDW ordering is observed. The Sb-Sb bond length within the square net is 3.0574(2) \AA. The Sb-Sb bond lengths in the square net are significantly longer than what would be expected in a localized and covalent Sb-Sb single bond, as it appears for example in elemental Sb (approximately 2.9\AA at 78 K)\cite{barrett_crystal_1963}. The elongated, symmetric bonding can be rationalized by ``hypervalent'' bonding, where a -1 charge on Sb is stabilized by electron delocalization across the entire net, as detailed by Papoian and Hoffmann. \cite{a2000hypervalent} The NdTe layer of the structure displays the aformentioned puckered rock salt structure, with Nd-Te bond length nearly parallel to \textit{ab} (within the unit cell) held constant at 3.1931(2) \AA, and bond length parallel to \textit{c} (across unit cells) held at 3.2788(6) \AA.

The structure of NdSb$_{0.48}$Te$_{1.37}$ displays a CDW distortion, as evidenced by the appearance of low intensity satellite peaks visible in the precession images of the diffraction patterns (Figure S2). The pattern can be indexed as an incommensurately modulated structure, with a modulation wavevector \textit{\textbf{q}}=0.18\textit{\textbf{b}}$^{\ast}$. This modulation wavelength corresponds to an approximate five-fold expansion of the unit cell along \textit{\textbf{b}}$^{\ast}$. Using the superspace approach to structure solution and refinement, NdSb$_{0.48}$Te$_{1.37}$ is found to crystallize in the $Pmmn$(0$\beta$0)00$s$ superspace group, and when visualized as a commensurate five-fold approximant, is isostructural to CeSb$_{0.51}$Te$_{1.40}$ and GdSb$_{0.46}$Te$_{1.48}$, where the (Sb,Te) square net is found to distort into a pattern of zigzag chains and more isolated atoms. This is congruent with our previous studies on the evolving CDWs observed in GdSb$_{x}$Te$_{2-x-\delta}$ and CeSb$_{x}$Te$_{2-x-\delta}$.  This same structure was found to give rise to an ideal nonsymmorphic Dirac semimetal band structure.\cite{lei2021band}

Using the Sb-Sb bond length in tetragonal NdSb$_{0.91}$Te$_{0.89}$ as a reference for an undistorted square net with fully delocalized bonding, we discuss the localization and symmetry breaking of the square net in NdSb$_{0.48}$Te$_{1.37}$. The square net breaks into a repeating pattern of parallel chains separated by lines of more isolated (Sb,Te) atoms (Fig 1d). Intra-chain Sb-Sb bonding varies from 2.92105(1) to 2.9593(1) \AA, and remains constant within each chain. The nearest distance from a chain atom to an isolated atom ranges from 3.1098(1) to 3.1706(1) \AA. Inter-chain atom distances are the largest, ranging from 3.2655(1) to 3.2872(1) \AA. The significant deviations from the hypervalent bond length found in the undistorted square-net imply a strong localization of bonding along the chain directions, with minimal bonding interactions between the chains. The chain-to-isolated atom distance sits in an intermediate range, larger than the expected value for delocalized bonding, but may carry more bonding interaction than inter-chain interactions.

The heavy distortion of the square net due to the CDW in NdSb$_{0.48}$Te$_{1.37}$ is contrasted with minimal changes to bond length within the NdTe layer. NdTe bond length nearly parallel to \textit{ab} held constant at 3.2122(1) \AA, and bond length parallel to \textit{c} is 3.2699(2) \AA. These bond lengths are quite similar to that in undistorted NdSb$_{0.91}$Te$_{0.89}$, with the deviation well explained by differences in the parent cell lattice parameters between the two. Figure S3a shows the modulation amplitude of bond distances in the NdTe structural unit and within the distorted square net. Nd-Te bonds do not vary more than 0.01 \AA, whereas the amplitude of bond modulations within the distorted square net vary as much as 0.4 \AA. (Figure S3b) This feature is common to all of the modulated LnSb$_{x}$Te$_{2-x-\delta}$ we have studied\cite{lei2019GdSbTe,singha2021evolving}, and consistent with our understanding of bonding in the square net. The CDW distortion is driven by electronic instability that arises from adding Te (and additional electrons) into the square net, breaking the bond delocalization. We consider the electrons involved in bonding in the NdTe layer to lie well below the Fermi energy, and thus will be unaffected by the CDW. This lack of structural interaction with the CDW will become interesting to consider in our discussion of the magnetic structure and its connection to the CDW.

\subsection{Magnetic measurements and specific heat}
\begin{figure}[hbt!]
  \includegraphics[width=\textwidth]{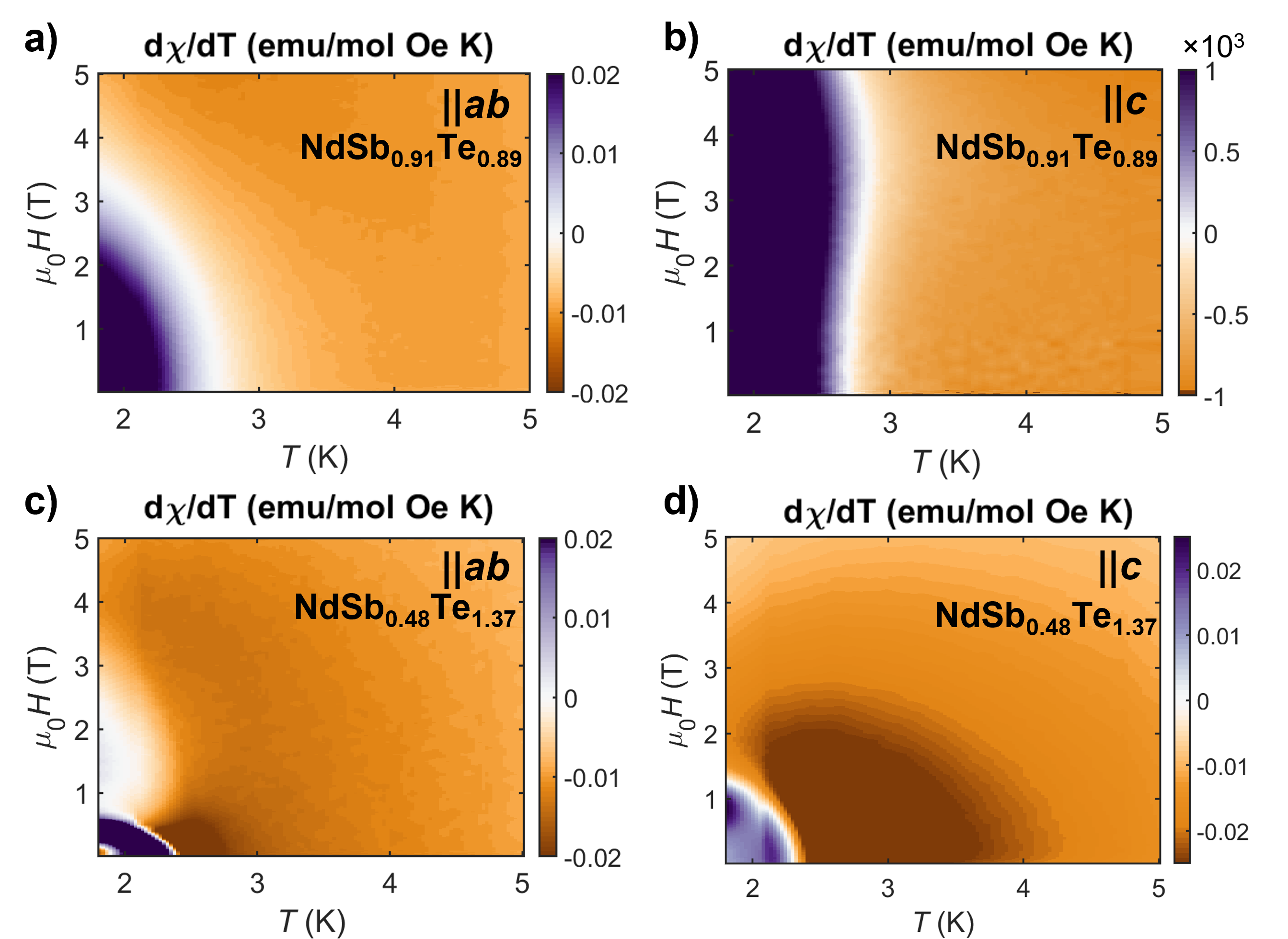}
  \caption{Magnetic phase diagrams (H vs T) of NdSb$_{0.91}$Te$_{0.89}$ and NdSb$_{0.48}$Te$_{1.37}$. Plots were generated from the first order derivative of $\chi$ over temperature (d$\chi$/dT).}
  \label{magphasediagrams}
  \centering
\end{figure}

In Figure 2, we present the magnetic phase diagrams of NdSb$_{0.91}$Te$_{0.89}$ (Fig 2. a, b)  and NdSb$_{0.48}$Te$_{1.37}$ (Fig 2. c, d), constructed by plotting the first-order derivative of $\chi$ ($T$) curves measured along the crystallographic \textit{ab} plane and \textit{c} axis. In the case of NdSb$_{0.91}$Te$_{0.89}$, a Neel temperature T$_{N}$ = 2.7 K is observed. A magnetic field of 3 T eventually suppresses the AFM order when applied along the \textit{ab}  plane(Fig 2a); the system presumably transitions to a fully spin-polarized state at this field. If the field is applied along the \textit{c} axis (Fig 2b), no metamagnetic transition is clearly visible from the variable field moment versus temperature measurements used to construct the phase diagram. Field-dependent magnetization curves show a subtle change in slope at 3 T in the derivative curve (Fig S4b), and a smaller magnetic moment at all fields compared to in-plane measurements (Fig S2a). Thus, the feature at 3 T may be an in-plane contribution due to imperfect alignment of the crystal in the applied field. Significantly different moments at the maximum applied field ($M_{7 T,\textit{ab}}\gtrsim 2M_{7 T,\textit{c}}$)  along in- and out-of-plane directions, further implies magnetic anisotropy (Fig S4a). These results are similar to previously reported magnetic measurements on tetragonal NdSb$_{x}$Te$_{2-x-\delta}$, where \textit{x} = 0.8.\cite{pandey_electronic_2020}

In NdSb$_{0.48}$Te$_{1.37}$, the observed T$_{N}$ is 2.3 K, lower than the tetragonal compound. This contrasts behavior observed in Ce and Gd members of LnSb$_{x}$Te$_{2-x-\delta}$. CeSb$_{x}$Te$_{2-x-\delta}$ displays an increase in T$_{N}$ with decreasing \textit{x}; GdSb$_{x}$Te$_{2-x-\delta}$ shows nonlinear behavior where T$_{N}$ increases to a maximum at \textit{x} = 0.48, and decreases at lower Sb compositions. \cite{singha2021evolving, lei2021magphase} Measurements with fields applied within the \textit{ab} plane (Figure 2c) detect a metamagnetic transition at $\sim$0.6 T, with a region of enhanced $\chi$ that persists to 3 T, where another metamagnetic transition occurs, presumably to a fully spin-polarized state. Both phase boundaries are corroborated by magnetization vs field measurements (Figure S4d) collected at 1.8 K. Measurements with fields applied along the \textit{c} axis (Figure 2d) indicate a pair of metamagnetic transitions at $\sim$0.7 and $\sim$1.2 T, which is also observed in the magnetization curves at 1.8 K as a shoulder in the derivative curve followed by a peak with increasing field. The presence of more metamagnetic transitions at lower critical fields in both directions, implies that spins are more readily polarized along either crystallographic direction in comparison to NdSb$_{0.91}$Te$_{0.89}$. Field-dependent magnetization also shows a smaller difference between in- and out-of-plane saturating moments in NdSb$_{0.48}$Te$_{1.37}$ (Fig S4c), altogether indicating that the CDW phase displays less magnetic anisotropy than its undistorted relative. In addition, the CDW phase exhibits more complex magnetism as compared with the tetragonal one, as evidenced by the multiple metamagnetic transitions.

The magnetic phase diagram of NdSb$_{0.48}$Te$_{1.37}$ can be compared to the isostructural CDW-containing relatives GdSb$_{0.46}$Te$_{1.48}$ and CeSb$_{0.51}$Te$_{1.40}$.\cite{lei2021magphase, singha2021evolving} GdSb$_{0.46}$Te$_{1.48}$ exhibits a significantly more complex phase diagram than its tetragonal family member GdSb$_{0.85}$Te$_{1.15}$\cite{lei2019GdSbTe}, along all measured directions, but particularly under fields applied along the \textit{ab} plane, a behavior also observed in NdSb$_{0.48}$Te$_{1.37}$. In the latter case the magnetic ordering temperature is much lower and so the various magnetic transitions may be more difficult to resolve, whereas the Gd compound shows multiple temperature-induced phase transitions. The observed multiple in-plane transitions contrast with CeSb$_{0.51}$Te$_{1.40}$, where no metamagnetic transitions are observed if a field is applied along \textit{ab}, but multiple are observed for fields along \textit{c}. This corresponds with the out-of-plane magnetic structure of its stoichiometric parent, CeSbTe.

\begin{figure}[hbt!]\includegraphics[width=0.75\textwidth,height=0.75\textheight,keepaspectratio]{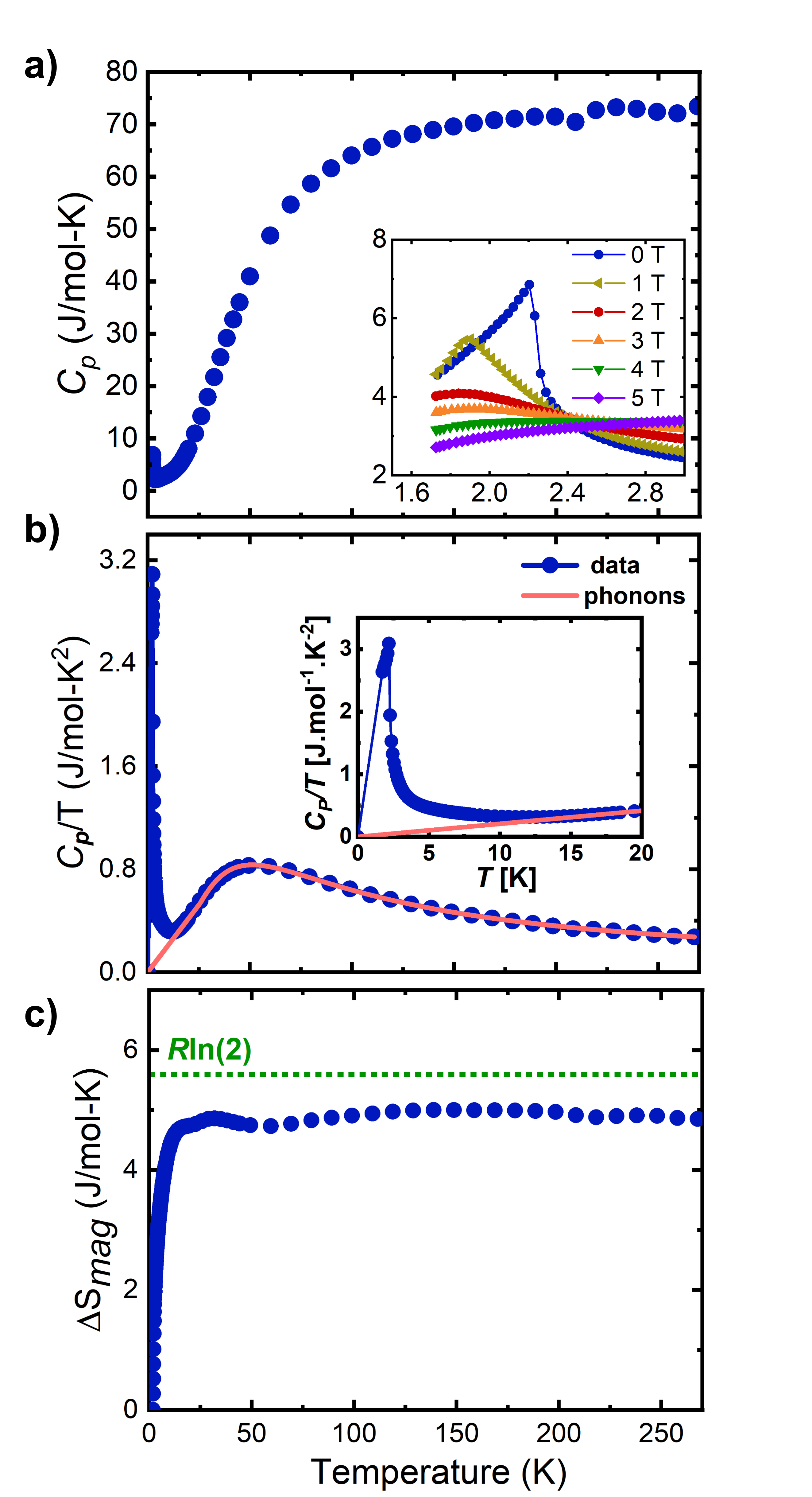}
  \caption{Heat capacity measurements. a) Temperature dependence of the heat capacity \textit{C$_{P}$} for NdSb$_{0.48}$Te$_{1.37}$ at variable applied fields. Inset focuses on low-temperature behavior at applied fields, highlighting the suppression of T$_{N}$ with applied field. b)Debye fit of \textit{C$_{P}$/T} with fitted phonon contribution shown. c) Magnetic entropy \textit{S$_{m}$} calculated from heat capacity data.}
  \label{heatcap}
  \centering
\end{figure}

To corroborate the observed T$_N$ and field dependence of the magnetic transitions in NdSb$_{0.48}$Te$_{1.37}$, we performed heat capacity measurements on a single crystal. At zero applied field, the observed T$_N$ is identical to that observed in the $\chi$ plots, and is suppressed with applied field in a manner consistent with them. The spin state of NdSb$_{0.48}$Te$_{1.37}$ can also be determined by integrating the magnetic component of the total heat capacity over temperature ($\frac{C_{P}}{T}$) to calculate the change in the magnetic entropy ($\Delta S_{mag}$). To determine the magnetic heat capacity, $C_{mag}$, the phonon contribution to heat capacity was subtracted using a single Debye mode as described in the equation shown below:

\begin{equation}
C_{D}({\theta_D,T}) = 9sR\left(\frac{T}{\theta_D}\right)^3\int\limits_0^\frac{\theta_D}{T}\frac{(\theta/T)^4e^{\theta/T}}{[e^{\theta/T}-1]^2}d \frac{\theta}{T}
\end{equation}

where $s$ is the oscillator strength, $R$ is the molar Boltzmann constant, and $\theta_{D}$ is the Debye temperature. The parameters extracted from the least-squares fit between $T$ = 25 - 265K is $s$ = 3.007(8) and $\theta_{D}$ = 182.8(4) K respectively. The total oscillator strength is in reasonable agreement with the total number of atoms per formula unit in NdSb$_{0.48}$Te$_{1.37}$. In the phonon fit, an Einstein mode was not incorporated because there was no $T_{max}$ in the $\frac{C_{P}}{T^3}$ plot. Thus, a single Debye mode was sufficient for the phonon fit. The $\Delta S_{mag}$ with respect to temperature is found to saturate at $\simeq$4.99 J/mol-K, slightly below what would be expected for a $S=\frac{1}{2}$ spin state ($R$ln2). The slight decrease in the $\Delta S_{mag}$ result from additional entropy release below $T$ = 2 K that is not captured by a linear extrapolation of $C_p/T$ $\rightarrow$ 0  at $T$ = 0 K. The presence of artifacts and magnetic entropy beyond the phonon subtraction region are viable possibilities; thus the heat capacity analysis had some limitations. Our results notably contrast with entropy observed by Pandey et. al. in NdSb$_{x}$Te$_{2-x-\delta}$, where \textit{x} = 0.8, which more closely indicates a $S = \frac{3}{2}$ state consistent with a free Nd$^{3+}$ ion. \cite{pandey_electronic_2020}

\subsection{Neutron diffraction and magnetic structures}
The magnetic ground state in zero applied magnetic field for both compounds has been investigated through powder neutron diffraction in the 5 -- 1.5 K temperature range. The paramagnetic data set of NdSb$_{0.94}$Te$_{0.92}$, collected at 5 K, is in good agreement with the structural model obtained on single crystal specimens (see Figure S5 and Table S3 and S4 in the supporting information for details). On cooling below $T_N\approx 2 K$, broad magnetic reflections are observed in the diffraction pattern, which indicates a magnetic correlation length significantly shorter than the nuclear one. The peaks can be indexed with the propagation vector k=(0 1/2 1/2). The possible magnetic isotropy subgroups consistent with both, the parent structure symmetry and the observed propagation vector, were obtained with the help of the ISODISTORT software.\cite{Campbell2006} The best agreement with the experimental data is obtained for the magnetic space group A$_{b}$2$_{1}$/c (P$_{c}$2$_{1}$/c, basis={(0,1,1),(-1,0,0),(0,-2,0)} origin=(1,1,0) in standard settings, see table S3 in the SI for details). The refined magnetic structure, together with the magnetic Rietveld plot, are shown in figure 4a and 4b. The spin ordering consists of AFM chains that run along the b axis, where the spins lie parallel to b and antiparallel to each other along the chain. A refined moment of 1.674(8)$\mu _{B}$ per Nd was obtained. The two AFM chains within a NdTe layer are shifted by half a unit cell in plane, due to the presence of the n-glide symmetry operation in the parent structure, which leads to one ferromagnetic ($J_1$) and one antiferromagnetic ($J_2$) Nd-Nd interaction. This geometrical frustration of the intra-chain interactions could be the reason of the very broad magnetic reflections observed in the neutron data. In fact, in VOCl, which is structurally closely related to tetragonal NdSb$_{0.94}$Te$_{0.92}$, a lattice distortion, which lifts this frustration, is observed at the magnetic ordering temperature.\cite{komarek_strong_2009} Since no such distortion is observed NdSb$_{0.94}$Te$_{0.92}$, the frustration between the intra-chain interaction remains, resulting in the broad magnetic Bragg peaks. In general, the observed magnetic structure is typical for PbFCl-type compounds, as VOCl adopts a very similar one. \cite{komarek_strong_2009} Still, it is different than that observed in CeSbTe, where spins point along the c axis and the AFM coupling doubles the unit cell.\cite{schoop2018tunable}

The paramagnetic diffraction pattern of modulated NdSb$_{0.48}$Te$_{1.37}$, collected at 5 K, is in agreement with the structural model obtained from the single crystal data (see Table S7-8 and Figure S6 in the SI for details). The refinement has been conducted in the $Pmnm(00\gamma)000$ superspace group and due to the small intensity of the modulation satellite peaks, only the first order positional modulation parameters have been refined. The obtained structural parameters are reported in the SI and the refined value of the modulation vector is $q_{CDW}$=(0,0.183(1),0) (with respect to the parent P4/nmm structure). On cooling below T$_N$=2.0(1)K, sharp magnetic reflections are observed in the diffraction pattern. These can be indexed with the propagation vector $q_{mag}$=(0,-0.408(1),0), the minus sign and the relation with the nuclear modulation vector will be discussed in detail later. The possible magnetic superspace groups consistent with the observed modulation vector and the tetragonal $P4/nmm$ parent structure have been calculated with the help of the ISOTROPY and ISODISTORT software.\cite{Campbell2006,Stokes_ISO} The best agreement with the experimental data has been achieved with the $Pmnm.1'(00\gamma)s00s$ superspace group and the obtained spin structure and the relative magnetic Rietveld plots are shown in Figure 4c,d (see table S9 and S10 in the SI for detail). The magnetic structure can be refined as an elliptical cycloid with a sinusoidal in plane amplitude that is $\approx3$ times larger than the out of plane cosinusoidal component. The rotation of the moments between adjacent unit cells along the propagation vector is of $\pi-\phi$ (the sign choice of the propagation vector component change the rotation angle from $\pi-\phi$ to $\pi+\phi$). Due to the presence of the $\{n|0000\}$ symmetry element in the superspace group, the two cycloids on the Nd atoms in the NdTe layer are rotating in opposite directions. This suggests that the magnetic interaction between these two positions will average to zero over the period of the magnetic structure. The solution shown in Figure 4 is the only model that is able to fit the observed intensity correctly; attempts to constrain a constant moment structure with a circular cycloid return worse fits of the data, as shown in Figure S7. The same is true for attempts using a positive propagation vector i.e. a cycloid with a rotation angle of $\pi+\phi$ (Figure S8). 

\begin{figure}[hbt!]
  \includegraphics[width=\textwidth]{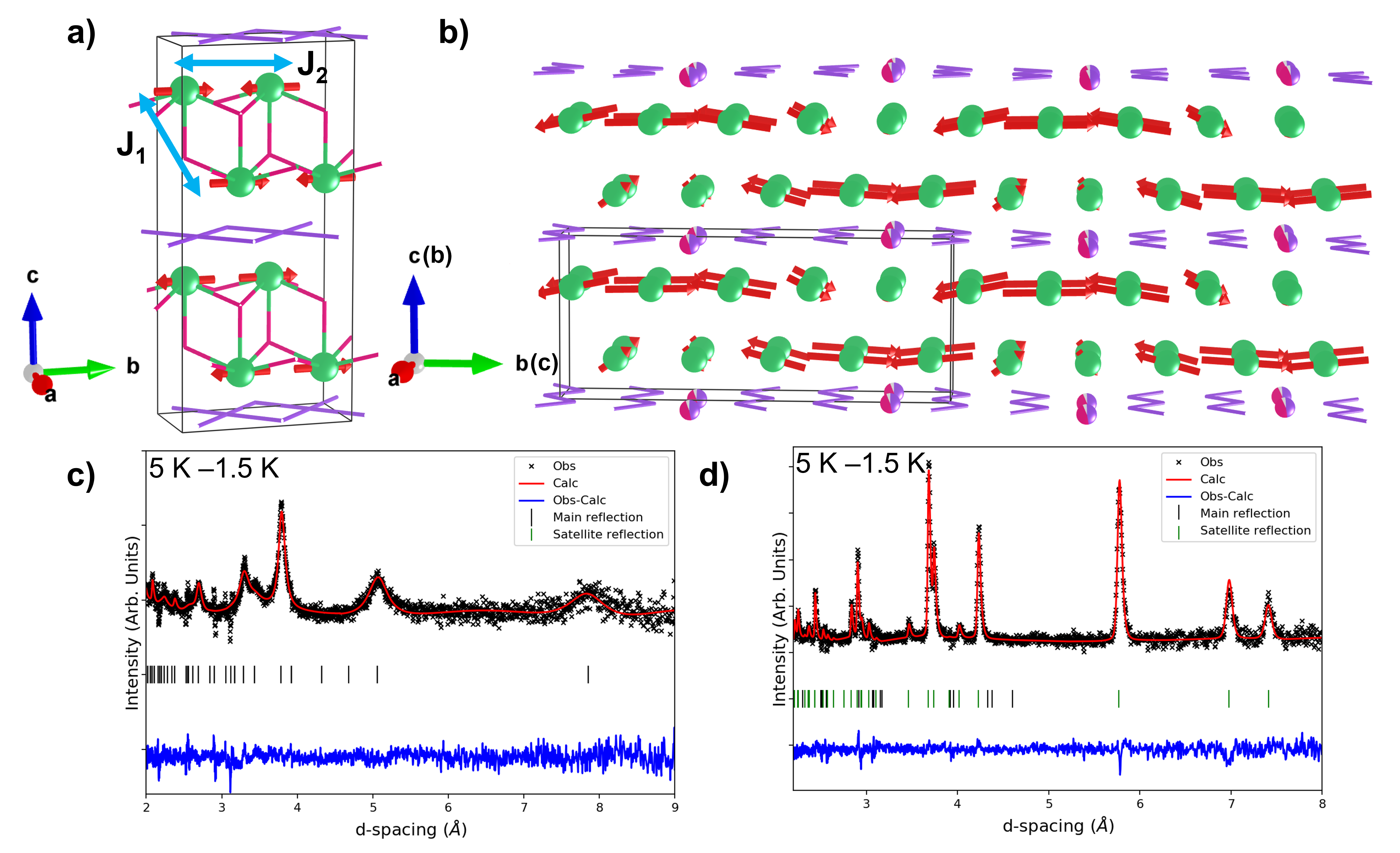}
  \caption{Magnetic structures determined by neutron diffraction. a) Magnetic structure of NdSb$_{0.94}$Te$_{0.92}$. b) Magnetic structure of NdSb$_{0.48}$Te$_{1.37}$, with exchange coupling terms $J_1$ and $J_2$ shown. In parenthesis, standard crystallographic axes used in Rietveld model are given. c) Magnetic Rietveld fit for NdSb$_{0.94}$Te$_{0.92}$, with magnetic scattering at 1.5 K plotted by subtraction of nuclear scattering collected at 5 K. d) Magnetic Rietveld fit for NdSb$_{0.48}$Te$_{1.37}$, with magnetic scattering at 1.5 K plotted by subtraction of nuclear scattering collected at 5 K. }
  \label{heatcap}
  \centering
\end{figure}

\subsection{Coupling between the CDW and the elliptical cycloid}

It is worth underlining the strong coupling between the observed elliptical magnetic structure and the charge density wave, which occurs above the magnetic transition. By observing the value of the modulation vectors, one can notice that they are related by $1+2\nu_2=\nu_1$ where $\nu_1$ and $\nu_2$ are the components of the CDW and cycloidal modulation vectors, respectively. Indeed, this relation can be rewritten, in vectorial form, as q$_{CDW}$+2q$_{mag}$=$(0\bar{1}0)$ which suggest that the nuclear modulation can be considered the second harmonic of the magnetic distortion, with the caveat that with respect to the parent $P4/nmm$ structure the two propagation vectors have opposite sign (the sign choice being a consequence of symmetry decomposition from the parent structure).

With respect to the parent $P4/nmm$ structure the nuclear modulation transforms as the $\Delta_3$ irreducible representation with the order parameter $(\delta,\delta^*;\eta,\eta^*)$, defined in a complex basis, and modulation vector $q_{CDW}=(0,\nu_1,0)$. On the other hand, the elliptical magnetic structure transform as the time-even irreducible representation $m\Delta_2$ with the order parameter $(\mu,\mu^*;\xi,\xi^*)$ and propagation vector $q_{CDW}=(0,\nu_2,0)$. By inspecting the transformation matrices of the generators of the $P4/nmm.1'$ space group for these two irreducible representations (reported in table S9), it is possible to write a third degree free energy invariant that couples these two distortions. The coupling term is $ \delta^*\mu^2+\delta\mu^{*2}+\eta^*\xi^2+\eta\xi^{*2}$ and it is allowed only if the components of the propagation vector of the two distortions verify the condition $1+2\nu_2=\nu_1$. This coupling term gives a natural explanation why the system decides to order magnetically with the observed propagation vector and the observed symmetry. In fact, by choosing the $m\Delta_2$ symmetry for the magnetic order parameter and the value of the magnetic propagation vector such that $1+2\nu_2=\nu_1$, the system lowers its free energy by coupling to the nuclear modulation already present above the magnetic transition. The coupling term also provides the system a mechanism to relieve the magnetic frustration as can be appreciated from the sharper magnetic peaks in NdSb$_{0.48}$Te$_{1.37}$ compared to NdSb$_{0.94}$Te$_{0.92}$. The coupling term alone does not justify the elliptical character of the cycloidal structure, which is likely the result of the competition between the magnetic anisotropy observed in the magnetization data and the magnetoelastic coupling term with the CDW. The energy gain offered by activating the coupling term could resolve the frustration of the Nd-Nd interactions by stabilizing the elliptical cycloid structure instead of the collinear structure observed in tetragonal NdSb$_{0.94}$Te$_{0.92}$. 

 To further confirm the coupling term, the diffraction data collected at 1.5 K were refined in the $Pmnm.1'(00\gamma)s00s$ magnetic superspace group, refining the magnetic modulation as first order harmonics and the nuclear modulation as second order harmonics. The results of the refinement are summarised in table S9 and S10 and the corresponding Rietveld plots are shown in Figure S9. As can be appreciated, the refined nuclear modulation is in good agreement with the refined structure at 5 K, as well as the single crystal data.

 \subsection{Discussion}

 Both charge- and spin-density waves are phenomena driven by Fermi surface nesting, and are frequently considered as competing phase transitions, as they both result in the opening of an energy gap to relieve the electronic instability of the nesting.\cite{gruner_density_1994} The coexistence of both in a system implies the coupling of charge and spin degrees of freedom, which has been observed in metallic Cr \cite{fawcett_spin-density-wave_1988,hu_real-space_2022}, FeTe$_{1-y}$\cite{enayat_real-space_2014}, and in layered nickelate La$_4$Ni$_3$O$_{10}$\cite{zhang_intertwined_2020}, and recently, KMn$_6$Bi$_5$\cite{bao_spin_2022}. In a majority of these cases, CDW and SDW onset coincide at the same transition temperature, frequently attributed to magnetoelastic coupling, and the CDW can be considered as a secondary order parameter induced by the SDW. \cite{toledano_landau_1987} In NdSb$_{0.48}$Te$_{1.37}$, the CDW is observed at room temperature, well above the appearance of the spin modulation. In this case, the spin modulation is caused by the RKKY interaction, where the ordering of localized Nd moments is mediated by the conduction electrons that come from the (Sb,Te) square net,
 which carry the information of the lattice modulation to the spins. Thus, the CDW plays a role in ``templating'' the resulting spin wave below T$_N$, as well as relieving the magnetic frustration apparent by comparison to NdSb$_{0.94}$Te$_{0.92}$. A similar ``templating'' effect has been observed in $CaMn_7O_{12}$ \cite{Johnson2016} in which the observed orbital density wave \cite{Perks2012} influence the exchange interactions stabilizing the intermediate temperature SDW, \cite{Johnson2016} and in $\zeta-Mn_2O_3$ where the commensurate orbital order drives the selection of the magnetic ground state.\cite{Khalyavin2018} The ``templating'' effect of the CDW and the subsequent relief of the exchange frustration explains why the temperature induced antiferromagnetic transitions occur more sharply in NdSb$_{0.48}$Te$_{1.37}$. The lower critical field for metamagnetic transitions, both in-plane (\textit{ab}) and in the out-of-plane direction (\textit{c}), agree well with the magnetic structure, since the elliptical cycloid lies in the \textit{bc} plane.
 
 The coupling term constructed for NdSb$_{0.48}$Te$_{1.37}$ also implies that for a CDW phase involving 2 arms of the star of \textbf{q}, the resultant magnetic structures will have the same number of arms involved. Given that multi-\textbf{q} magnetic states are known to give rise skyrmion phases and other related spin textures, it is possible that CDW can be used as a tool to access them via RKKY interactions in addition to the more-common Dzyaloshinskii–Moriya interaction which drives their presence in other systems.  Indeed, it has been shown that densely packed skyrmion lattices can exist in centrosymmetric crystals, where the skyrmions are believed to originate from frustrated RKKY interactions. \cite{takagi_square_2022, hirschberger_skyrmion_2019, khanh_nanometric_2020, kurumaji_skyrmion_2019} Recent powder neutron diffraction on stoichiometric HoSbTe and TbSbTe, tetragonal members of  \textit{Ln}Sb$_{x}$Te$_{2-x-\delta}$, reveal complex, multi-\textbf{q} magnetic structures, requiring multiple commensurate propagation vectors in the ground state, which transform to incommensurate upon warming while still below T$_{N}$.\cite{plokhikh_competing_2022} Exploring the synthesis of their CDW-containing derivatives may hold potential to access such skyrmion lattices. In previously mentioned GdSb$_{0.46}$Te$_{1.48}$, a low-field region of the magnetic phase diagram in the out-of-plane direction shows possible signatures of an antiferromagnetic skyrmion phase by magnetoentropic analysis. The multiple low-field transitions present in the out-of-plane magnetization of NdSb$_{0.48}$Te$_{1.37}$ may bear relation to the magnetic phases observed in GdSb$_{0.46}$Te$_{1.48}$, also hinting to the role that the CDW may play therein. A recent single-crystal neutron diffraction study on single-q, near-6-fold CDW phase GdSb$_{0.71}$Te$_{1.22}$ reports the appearance of two magnetic phases which arise subsequently below T$_N$ and below the final transition at 5 K, respectively. \cite{plokhikh_magnetic_2023} The multi-\textbf{q} CDW phase CeSb$_{0.10}$Te$_{1.79}$ shows complex out-of-plane metamagnetic transitions in the form of a ``devil's staircase''\cite{singha2021evolving} in the out-of-plane magnetization, where it is also suggested that the CDW couples to a type of spin wave. However, in this system, the CDW wavevectors contain out-of-plane components, which may allow interaction with the expected out-of-plane spin structure.

\section{Conclusion}
In conclusion, we have investigated the crystal structure, magnetic properties, and magnetic structure of the undistorted square-net TSM NdSb$_{0.94}$Te$_{0.92}$ and its CDW-distorted relative NdSb$_{0.48}$Te$_{1.37}$ to understand the interactions between the charge density wave and magnetism in \textit{Ln}Sb$_{x}$Te$_{2-x-\delta}$ systems. NdSb$_{0.94}$Te$_{0.92}$ is isostructural to other stoichiometric \textit{Ln}SbTe, where delocalized bonding stabilizes the Sb square net. NdSb$_{0.48}$Te$_{1.37}$ exhibits a CDW distortion that breaks the square net into zig-zag chains with isolated atoms, isostructural to other \textit{Ln}Sb$_{x}$Te$_{2-x-\delta}$ with similar \textit{x} and nearly commensurate (\textbf{q}=0.18\textbf{b$^{\ast}$}).  While NdSb$_{0.94}$Te$_{0.92}$ behaves as a typical collinear antiferromagnet with a T$_N$ of 2.7 K , NdSb$_{0.48}$Te$_{1.37}$ displays additional complex magnetism, with a lower T$_N$ of 2.3 K and additional metamagnetic transitions. NdSb$_{0.48}$Te$_{1.37}$ has an elliptical cycloid magnetic structure with a propagation vector \textit{\textbf{q}}=-0.41\textit{\textbf{b}}$^{\ast}$. The near-integer multiple of the CDW vector provides a rare example of CDW inducing complex spin order through coupling of propagation vectors, as well as the most direct evidence of CDW coupling with magnetism  in \textit{Ln}Sb$_{x}$Te$_{2-x-\delta}$ to date. This is also the first example of a  magnetic structure of a CDW phase within the \textit{Ln}Sb$_{x}$Te$_{2-x-\delta}$ family. The existence of the CDW above the magnetic transition temperature suggests that it plays a role in ``templating'' the spin interactions between \textit{Ln} atoms via RKKY interaction. Furthermore the complexity of the magnetism and the richness of CDW phases accessible in \textit{Ln}Sb$_{x}$Te$_{2-x-\delta}$ phases, suggest that it is possible to achieve RKKY interaction-induced skrymions in these materials.

\begin{acknowledgements}

This work is supported by an NSF CAREER grant (DMR-2144295) to LMS. Additional support was provided by the Air Force office of Scientific Research under award number FA9550-20-1-0246. This work was further supported by the Gordon and Betty Moore Foundation (EPiQS Synthesis Award) through grant GBMF9064, the David and Lucile Packard Foundation, the Alfred P. Solan foundation, and the Arnold and Mabel Beckman foundation through a BYI grant awarded to LMS and an AOB postdoctoral fellowship awarded to JFK. TB acknowledges support from NSF-MRSEC through the Princeton Center 
for Complex Materials NSF-DMR-2011750. THS is grateful to Marceline Martineau for helpful discussions. F.O. and P.M. are grateful to Dr. Dmitry Khalyavin for the very fruitful discussion. For the purpose of open access, the author has applied a creative commons attribution (CC BY) licence to any author accepted manuscript version arising. 

\end{acknowledgements}


%

\end{document}